\documentstyle[psfig,12pt]{article} % gallery format

\begin{document}
\title{Using Entropy-Based Methods to
  Study \\General Constrained Parameter Optimization Problems\\
\vskip 0.5cm
\small{M. Argollo de Menezes$^{1}$ and A. R. Lima$^{2}$
\footnote{Electronic Addresses: marcio@if.uff.br
    , arlima@pmmh.espci.fr}\\
  $1)$ Instituto de F\'{\i}sica, Universidade Federal Fluminense \\ 
  Av.  Litor\^anea 24210-340, Niter\'oi, RJ, Brazil
    , marcio@if.uff.br\\
  $2)$ Laboratoire de Physique et M\'echanique des Milieux
  H\'et\'erog\`enes, ESPCI Paris \\ 10 rue Vauquelin, 75231 Paris
  Cedex 05, France}\\
}
                                %
%\date{Version: 27/04/2001 - Printed: \today}
%
%
\maketitle

\begin{abstract}
  In this letter we propose the use of physics techniques for entropy
  determination on constrained parameter optimization problems. The main
  feature of such techniques, the construction of an unbiased walk on energy
  space, suggests their use on the quest for optimal solutions of an
  optimization problem.  Moreover, the entropy, and its associated density of
  states, give us information concerning the feasibility of solutions.
\end{abstract}

%{{\bf PACS:}  89.80+h, 02.70.Lq}

%89.80.+h Computer science and technology
%02.70.Lq Monte Carlo and statistical methods

%-------------------------------------------------------------------------
%\section{ Introduction }
%\label{sec:intro}
%-------------------------------------------------------------------------
Statistical physicists are constantly developing new computational and
theoretical tools to unravel the complex behavior of systems composed by many
simple, interacting units. In the last years, many works ~\cite{books} have
showed us that one could ``translate'' a variety of problems, ranging from
biology to economy, to a physicist's language. One such example is the work of
Ros\'e et.  al., who extended the concept of ``density of states'' to complex
optimization problems\cite{roseetal96,rose97}.

This paper follows the same guidelines: we propose the use of methods which
obtain the entropy of physical systems to study constrained parameter
optimization problems. These include important tasks, such as production
planning, decision problems, structural optimization \cite{michalewiczetal96}
and design of semiconductor quantum devices \cite{goldonirossi00}.  From the
operational point of view these problems are usually analytically untractable
and numerically hard to solve, mainly due to the intricate shape of the space
of feasible solutions.

A general constrained optimization problem can be formulated as follows:

\begin{eqnarray}
\label{eq:problem}
\mbox{Minimize (or Maximize)~~~}& & E(\vec{x})\\
\mbox{Subject to~~~} & & g_i(\vec{x}) \leq 0, \mbox{~~i=1,...,q}\nonumber\\
                     & & h_i(\vec{x}) = 0,  \mbox{~~i=q+1,...,m}\nonumber
\end{eqnarray} 
where $\vec{x}=(x_1, x_2, x_3, ..., x_n)$ is an $n$-component vector,
$g_i(\vec{x})$ are $q$ inequality constraints and $h_i(\vec{x})$ are
$m-q$ equality constraints. The functions $E(\vec{x})$, $g_i(\vec{x})$
and $h_i(\vec{x})$ can be either linear or nonlinear, continuous or
discontinuous. {\em Feasible solutions} can be defined as the set $A(\{\vec{x}\})$ of
vectors which satisfy all the constraints, and {\em optimal solutions}
as the subset $B\in A$ of the feasible ones which minimize (maximize) the cost
$E$.

Although there are many well-established heuristic algorithms for such
problems, like evolutionary algorithms
\cite{goldonirossi00,michalewiczschoenauer96} and Constrained
Simulated Annealing\cite{wahchen00,wahwang00}, none of them give, in
a simple way, any information about the structure of the problem.

The ``translation'' of the abovementioned problem to a physicist's language
proceeds as follows: the variables $x_i$ of the optimization problem
(\ref{eq:problem}) are identified with the simple interacting units (pointed
out in the first paragraph) and the vector $\vec{x}$ with a physical state.
$E(\vec{x})$ is the energy (or cost) of the state $\vec{x}$. The density of
states is here identified with the ratio between the number of solutions with
a given cost $E$ and the total number of solutions.  Mathematically it reads
\begin{equation}
  g(E) = \frac{\sum_{\vec{x}_f}{ \delta_{E(\vec{x}_f) - E}}}{\sum_{\vec{x}_f}{1}}
\end{equation}
where the summation runs over all feasible solutions. The entropy $S(E)$, in
units of $k_B$, is defined as $S(E)=\ln g(E)$.

As pointed by H.  Ros\'e \cite{roseetal96}, independently of the complexity of
the space of solutions, the density of states is a direct measure which
reflects how sparsely states with a given quality $E$ are distributed.  Thus,
from the density of states (or, equivalently, the entropy $S(E)$) we can
estimate whether solutions with a given cost $E$ represent global optimal
values or if further optimization could lead to better results with reasonable
computational effort\cite{roseetal96}.  Unfortunately, it is not a simple task
to obtain the entropy or to estimate it from traditional simulations
(Metropolis importance sampling, microcanonical simulations, etc...)
\cite{newman99,landaubinder00}.  Over the past few years, many efficient
algorithms have been developed aiming at an efficient calculation of this
quantity. Those include multicanonical methods~\cite{berg91}, the ``Broad
Histogram Method'' ~\cite{pmco96,pmco98a} 
and, more recently, the ``Multiple Range Random Walk'' algorithm (MRRW)
\cite{wanglandau00}.  Since the latter seems to be the most simple to
implement and easiest to generalize \cite{wanglandau00,huller00}, we adopted
it as a calculational tool to study optimization problems on this letter, and
before applying it to our problem, let us briefly review its most important
points. 

The main idea of the MRRW algorithm is to obtain the density of states (DOS)
recursively. For a good resolution on the DOS, one must visit the energy axis
on a non-biased way, what can be done by performing a random walk in energy
space with the probability of visiting a state $\vec{x}$ being proportional to
the reciprocal of the density of states, $1/g(E)$. Since $g(E)$ is not known
{\it a priori}, it is set to $g(E)=1$ for all $E$ at the beginning of the
random walk, and then configurations are randomly chosen. If a given choice
changes the energy of the system from $E_1$ to $E_2$, the new configuration is
accepted according to

\begin{equation}
\label{eq:transition}
p(E_1\to E_2)=\mbox{~min}\left(\frac{g(E_1)}{g(E_2)},1\right).
\end{equation}
Then, after the energy level $E$ is visited, $g(E)$ is updated on a
multiplicative fashion, $g(E)\to g(E)\times f$, and the histogram of visits
$H(E)$ on an additive way, $H(E)\to H(E)+1$. When fluctuations on
$H(E)$ are sufficiently small (the magnitude of $f$ is related to the
``flatness'' of the histogram), the multiplicative factor $f$ is
decreased to $f^{\prime}=\sqrt{f}$. This sequence is repeated until a
predefined value of f is achieved (for more details, we refer to the original
paper ~\cite{wanglandau00}.

In principle, this prescription for a ``flat histogram'' assures that all
energy levels are visited equally and, in practice, one expects that the
extreme values of the accessible energies will be visited with the same
probability as any other energy level. Since on optimization problems one is
interested in extremal values of a cost function (or energy) which
characterizes the best solution of a given problem, is natural to think that
``flat histogram'' algorithms are good candidates for solving optimization
tasks, as we will show on the rest of the paper.

A great advantage of these methods over traditional techniques is that there
are no restrictions on how one should make changes to the state ${\vec{x}}$,
as long as the new state sampled is accepted according to the transitional
probability (\ref{eq:transition}).  So, this choice can always be made such
that it handles efficiently the constraints of the problem ~\cite{additional}.

Let us take as a simple test case a parabola $E(x)=x^2$ on the interval $0
\leq x \leq 1$, with $x$ being a random variable. From the analysis of
functions of random variables ~\cite{reif}, one can obtain the probability
distribution function $P(E)dE$, from which the entropy $S(E)$ can determined.
In this simple case it reads $S(E)=-\ln{E}/2$. To recover this entropy with
the MRRW algorithm we simply need to initialize our ``system'' with an initial
``state'', for instance $x=0.5$, then choose randomly another ``state''
$x^{\prime}$. The latter is accepted according to the transitional probability
(\ref{eq:transition}), and then histograms are updated. After accomplishing
all steps of the method, ~\cite{recipe-for-mrrw}, we compare our numerical
estimate with the analytical result (fig. $1$) and find a root mean square
deviation of order $10^{-2}$.
\begin{figure}[!h]
\label{fig:e}
  \vspace{1cm}
{\centerline{\psfig{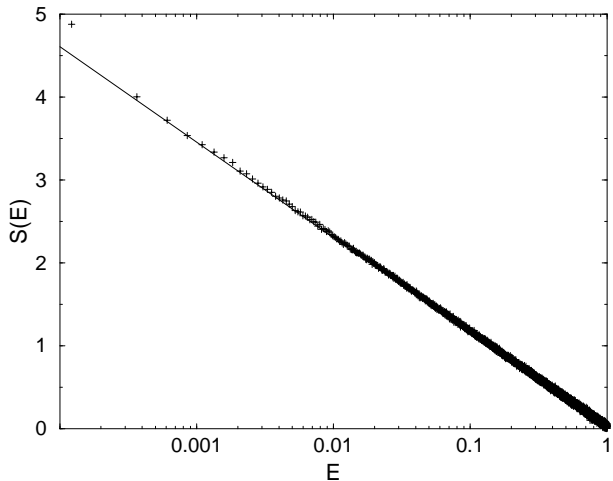}}}
\caption{Comparison between numerical and analytical results
for the entropy of the function $E=x^2$ of a random variable $0<x<1$.}
\end{figure}

Once we verified the correctness of the method, let us focus on 
harder problems:

\vspace{1cm}
$1)$ Find the maximum of
\begin{equation}
  E_1(\vec{x}) =  \left| \frac{\sum_{i=1}^n{\cos^4(x_i)} -
      2\prod_{i=1}^{n}{\cos^2(x_i)}}{\sqrt{\sum_{i=1}^n{i x_i^2}}}\right|
\end{equation}
subject to $\prod_{i=1}^{n}{x_i} \geq 0.75$, $\sum_{i=1}^n{x_i} \leq
7.5n$ and $0 \leq x_i \leq 10$.

\vspace{1cm}
$2)$ Find the maximum of 
\begin{equation}
  E_2(\vec{x}) = \left(\sqrt{n}\right)^n \prod_{i=1}^n{x_i}
\end{equation}
subject to $\sum_{i=1}^{n}{x_i^2} = 1.0$. 

These are well-known test functions for optimization
algorithms~\cite{micha-book}, whereas the optimal solution for the first
problem is not known exactly, while for the second one the maximum is at
$E=1$.

Following the observation that very often the global solution of many
constrained numerical optimization problems lies on the boundary of the
feasible region \cite{michalewiczschoenauer96}, we fix the search to the
``edge of feasibility''. For instance, in the first example we fix the
condition $\prod x_i=0.75$ and perform simultaneous changes to two variables
$x_i$ and $x_j$ at each change (that is, modify $x_i$ by a random
variable $0<q<2$ and divide $x_j$ by the same $q$). Doing so, we obtain the
entropy of the former problem, which is depicted in figure ($2$) for the case
$n=20$.

\vspace{1cm}
\begin{table}[!h]
  \begin{center}
    \begin{tabular}{|c|c|c|c|c|} 
      \hline
      Problem & n  & MRRW & EA \protect \cite{michalewiczetal96}& CSA \protect
      \cite{wahchen00,wahwang00} \\
      \hline
      $E_1$ & 20  & 0.803587 & 0.803553 & 0.803619 \\
      $E_1$ & 50  & 0.835131 & 0.833194 &        - \\
      $E_1$ & 100 & 0.845388 &        - &        - \\
      $E_2$ & 20  & 0.999881 & 0.999866 &      1.0 \\
      \hline
    \end{tabular}
    \caption{Comparison between numerical results obtained with multiple range
      random walks (MRRW), evolutionary algorithms (EA) and constrained
      simulated annealing (CSA).}
  \end{center}
\end{table}

On table $I$ we compare our results with those obtained with evolutionary
algorithms (EA) \cite{michalewiczschoenauer96} and constrained simulated
annealing (CSA) \cite{wahchen00,wahwang00}. We spent approximately $10^6$ MC
steps to obtain the entropies for both problems with a nice resolution (which
corresponds to $222$ seconds for problem $1$ with $n=20$ and $77$ seconds for
problem $2$, both running on a pentium II $400$ MHz), while the optimal
solutions are obtained in approximately $10$ times less MC steps. These
results were obtained from only one simulation, and could, in principle, be
improved by increasing the number of MC steps or by employing a steepest
descent method about the minimum found ~\cite{lima01}.  Our numerical
estimates are always better than those obtained with genetic algorithms, and
can get as good as constrained simulated annealing ones with this additional
(local) search~\cite{lima01}.

\begin{figure}[!h]
\label{fig:e1}
  \vspace{1cm}
{\centerline{\psfig{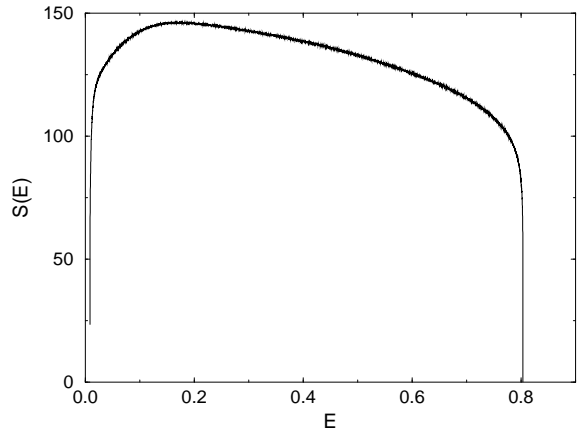}}}
\caption{Entropy as a function of the energy $E=E_1$ when
  $\prod_{i=1}^{n}x_i=0.75,~n=20$.}
\end{figure}

As a final test, we apply the MRRW algorithm to the traveling salesman problem
(TSP), an archetypal problem on computer science and one of the six basic
NP-complete problems\cite{garey79}. It can be formulated as the quest for the
shortest path which connects $N$ cities displaced on a plane map.  We try to
find optimal tours for two particular instances of the problem, known as ATT48
and kroA100, for which there are known solutions, mathematically proven to be
optimal \cite{tsplib}. In order to visit the space of states (the space of
feasible solutions) efficiently, we use the Lin Kerninghan algorithm
~\cite{lin-kern} to sample between valid tours.
We were able to find optimal tours with approximately $10^6$ MC steps in both
instances. On figure ($3$) we show the entropy of the kroA100 instance with
the optimal tour on the inset. From the entropy, one can directly assess how
difficult it is to find tours with a specified ``quality'', or perimeter $P$
~\cite{moreresults}.

\begin{figure}[htbp]
\label{fig:tsp1}
\vspace{1cm}
{\centerline{\psfig{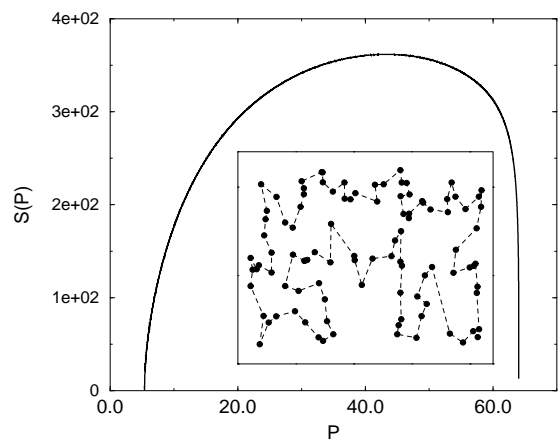}}}
\caption{Entropy of the kroA100 instance. With the MRRW algorithm we were
    able to find the tour with shortest perimeter $P$, which is exposed on the
    inset.}
\end{figure}

The aim of this paper was to propose the use of a simple and efficient
algorithm for entropy determination of physical systems on constrained
parameter optimization problems. By performing a biased walk on configuration
space, one is able to obtain an unbiased random walk on energy space and then
to reconstruct with good precision the density of states. This is the property
which makes this ``flat histogram'' algorithm a good tool for optimization
tasks, since the extreme energies of the system are visited at the same
frequency as other energies during the visitation scheme of the algorithm. As
a result, we were able to obtain, with reduced computational efforts,
satisfactory results for classical optimization problems.  As an additional
feature, we gain insight into the complexity of the problem by means of its
associated entropy.

\section*{Acknowledgments}

MAM has benefited from fruitful discussions with T.J.P.~Penna, while ARL thanks
L. Trujillo for helpful comments and discussions. We acknowledge financial
support from CNPq and CAPES (Brazilian agencies).


\begin{thebibliography}{99}
\bibitem{books} Per Bak, {\it How nature works}, Oxford Univ. Press (1997);
 S. M. Moss de Oliveira, P. M. C. de Oliveira and
  D.Stauffer, {\it Evolution, Money, Wars and Computers},
  Teubner-Stuttgart (1999);
  H. Levy, M. Levy and S. Solomon, {\it Microscopic Simulation of Financial
 Markets}, Academic Press (2000).

\bibitem{roseetal96}
H. Ros{\'e}, W. Ebeling, and T. Asselmeyer,  in {\em Parallel Problem Solving
  from Nature, Proc. of the Int. Conference on Evolutionary Computation},
  edited by H.~M. Voigt, W. Ebeling, I. Rechenberg, and H.~P. Schwefel
  (Springer, Berlin, 1996), No.~IV, pp.\ 208--217.

\bibitem{rose97}
H. Ros{\'e},  in {\em Proceedings of the International Conference on Complex
  Systems}, New England Complex Systems Institute, Y. Bar-Yam (ed.)
  (Nashua, NH, 1997).

\bibitem{michalewiczetal96}
Z. Michalewicz, D. Dasgupta, R. Le, and M. Schoenauer, Computers and Industrial
  Engineering Journal {\bf 30},  851 (1996).

\bibitem{goldonirossi00}
G. Goldoni and F. Rossi, cond-mat/0011404.

\bibitem{michalewiczschoenauer96}
Z. Michalewicz and M. Schoenauer, Evolutionary Computation {\bf 4},  1  (1996).

\bibitem{wahchen00}
B.W. Wah and Y. Chen,  in {\em Sixth International Conference on Principles
  and Practice of Constraint Programming} (Springer-Verlag, Berlin, 2000).

\bibitem{wahwang00}
B.W. Wah and T. Wang,  in {\em Sixth International Conference on Principles
  and Practice of Constraint Programming} (Springer-Verlag, Berlin, 2000).

\bibitem{newman99}
M.E.J. Newman and G.T. Barkema, {\it Monte Carlo Methods in Statistical
  Physics}, Oxford University Press (1999).

\bibitem{landaubinder00}
D.P. Landau and K. Binder, {\em A Guide to Monte Carlo Methods in Statistical
  Physics}, Cambride University Press, (2000).

\bibitem{berg91} B.A. Berg and T. Neuhaus, {\em Phys. Lett.}
{\bf B267}, 249 (1991); \\
B.A. Berg {\em Int. J. Mod. Phys.} {\bf C3}, 1083 (1992).

\bibitem{pmco96} P.M.C. de Oliveira, T.J.P. Penna and H.J. Herrmann,
  {\em Braz. J. Phys.} {\bf 26}, 677 (1996); cond-mat/9610041.
 
\bibitem{pmco98a} P.M.C. de Oliveira, T.J.P. Penna and H.J. Herrmann,
  {\em Eur. Phys. J.} {\bf B1}, 205 (1998).

\bibitem{wanglandau00}
F. Wang and D. P. Landau, Phys. Rev. Lett. 86, 2050 (2001); cond-mat/0011174.

\bibitem{huller00}
A. H\"uller and M. Pleiming, cond-mat/0011379.

\bibitem{additional}
 This method is also suitable for
parallelization, since one can divide the energy axis on $W$ windows, which
can be visited by independent random walkers on parallel on $W$ different
processors. Other strategy could be that of $W$ independent random walkers
along the whole energy axis, each using the same entropy $S(E)$ for the
transitional probabilities.

\bibitem{reif}
F. Reif, {\it Introduction to Statistical Physics}, Mc Graw Hill,
Int. Edition,  (1985).

\bibitem{recipe-for-mrrw}
For the implementation of the MRRW we start the quantity $f$ (see
  \cite{wanglandau00}) as $f=\exp(5.0)$. Its value is recalculated using
  $f_{\rm new}=\sqrt{f}$ after each $50000$ tries to change the value of all
  variables $x_i$. This process is repeated $15$ times.
  

\bibitem{micha-book} How to solve it: Modern Heruristics" Z. Michalewics and
  D. B. Fogel Springer, Berlin 2000

\bibitem{lima01}
A.~R. Lima, M.~A. de~Menezes, and L. Trujillo, to be published.

\bibitem{garey79}
M.~R. Garey and D.~S. Johnson, {\it Computers and Intractability: A guide to
  the theory of NP-Completeness}, Freeman (1979).

\bibitem{tsplib}
http://www.iwr.uni-heidelberg.de/iwr/comopt/software/TSPLIB95/

\bibitem{lin-kern} S. Lin and W. Kerninghan, Operations Research, Vol. 21, pp.
  498--516, 1973.

\bibitem{moreresults} We have tested the algorithm on a large number of
instances which are reported in \protect\cite{tsplib}. Extended results will
be presented in \protect\cite{lima01}.

\end{thebibliography}
\end{document}